\documentclass[fp,twocolumn]{jpsj3}
\usepackage{txfonts}
\usepackage[dvipdfmx]{graphicx}
\usepackage{color}
\usepackage{dcolumn}
\usepackage{bm}
\usepackage{amsmath,amssymb}
\usepackage{ulem}

\usepackage{color}
\ifx\color@rgb\@undefined\else
  \definecolor{kw1}{rgb}{0,0,1}
  \definecolor{kw2}{rgb}{0.3,1,0.5}
  \definecolor{kw3}{rgb}{1,0,0}

\title{Coexistence of Two Components in Magnetic Excitations of La$_{2-x}$Sr$_x$CuO$_4$ \\($x$ = 0.10 and 0.16)}

\author{Kentaro Sato$^1$, Kazuhiko Ikeuchi$^2$, Ryoichi Kajimoto$^3$, Shuichi Wakimoto$^4$, Masatoshi Arai$^5$, and Masaki Fujita$^6$\thanks{fujita@imr.tohoku.ac.jp}}

\inst{$^1$Department of Physics, Tohoku University, Aoba, Sendai 980-8578, Japan \\
$^2$Research Center for Neutron Science and Technology, Comprehensive Research Organization for Science and Society, Tokai, Ibaraki 319-1106, Japan\\
$^3$Materials and Life Science Division, J-PARC Center, Japan Atomic Energy Agency,  Tokai, Ibaraki 319-1195, Japan\\
$^4$Materials Sciences Research Center, Japan Atomic Energy Agency, Tokai, Ibaraki 319-1195, Japan\\
$^5$European Spallation Source ERIC, P.O. Box 176, SE-221 00 Lund, Sweden\\
$^6$Institute for Materials Research, Tohoku University, Katahira, Sendai 980-8577, Japan \\
}

\abst{
To elucidate the spin dynamics of La$_{2-x}$Sr$_x$CuO$_4$, which couples with the charge degree of freedom, the spin excitations spanning the characteristic energy ($E_{\rm cross}$ $\sim$35--40 meV) are investigated for underdoped $x$ = 0.10 and optimally doped (OP) $x$ = 0.16 through inelastic neutron scattering measurements. 
Analysis based on two-component picture of high-quality data clarified the possible coexistence of uprightly standing incommensurate (IC) excitations with gapless commensurate (C) excitations in the energy-momentum space. The IC component weakens the intensity toward $E_{\rm cross}$ with increasing energy transfer ($\hbar\omega$), whereas the C component strengthens at high-$\hbar\omega$ regions. The analysis results imply that the superposition of two components with a particular intensity balance forms an hourglass-shaped excitation in appearance. 
Furthermore, the temperature dependence of the intensity of each component exhibits different behaviors; the IC component disappears near the pseudo-gap temperature ($T$*) upon warming, whereas the C component is robust against temperature. These results suggest the itinerant electron spin nature and the localized magnetism of the parent La$_2$CuO$_4$ in IC and C excitations, respectively, similar to the spin excitations in the OP YBa$_2$Cu$_3$O$_{6+\delta}$ and HgBa$_2$CuO$_{4+\delta}$ with higher superconducting-transition temperatures. 
}

\begin{document}

\maketitle

\section{Introduction}
Spin dynamics in doped lamellar cuprate oxide is a fundamental issue in solid-state physics. Hole doping modifies the spin--wave excitation in Mott insulators~\cite{Coldea2001, Headings2010} to unconventional hourglass-shaped excitations in the superconducting (SC) phase of single-layer La$_{2-x}$(Sr, Ba)$_{x}$CuO$_4$ (La214)~\cite{Tranquada2004, Vignolle2007, Lipscombe2009} and double-layer YBa$_2$Cu$_3$O$_{6+\delta}$ (Y123)~\cite{Arai1999, Hayden2004}. Hourglass excitations are composed of four low-energy incommensurate (IC) peaks at (0.5$\pm\delta$,0.5)/(0.5,0.5$\pm\delta$) below the waist energy ($E_{\rm cross}$) and outwardly dispersive high-energy excitations above $E_{\rm cross}$. ~\cite{Tranquada2004, Vignolle2007, Lipscombe2009} Inelastic neutron scattering (INS) measurements on the optimally doped (OP) Y123 (SC transition temperature $T_{\rm c}$ = 93 K) indicated particle--hole pair excitations below $\sim$$E_{\rm cross}$ and local spin dynamics in higher energy transfers ($\hbar\omega$) region~\cite{Reznik2008}. 
Similarly, the {\it itinerant} spin nature in the gapped downward branch has been clarified for a nearly OP single-layer HgBa$_2$CuO$_{4+\delta}$ (Hg1201 $T_{\rm c}$ = 88K)~\cite{Chan2016}. Hence, the origin of the magnetic response would be common in OP cuprates with relatively high-$T_{\rm c}$, irrespective of the number of layers in the unit cell, i.e., the crystal structure. 

Meanwhile, in the La214 system, the origin of entire magnetic excitations are discussed in terms of {\it localized} spin and charge stripes because the one-dimensionally aligned spin and charge domains are statically stabilized around $x$ = 1/8~\cite{Tranquada1995}. Several theoretical models, including not only the stripe (ladder) model~\cite{Vojta2004, Tohyama2018},  but also the spiral~\cite{Kharkov2018}, Lifshitz spin liquid~\cite{Kharkov2019}, and Fermi surface nesting models~\cite{Norman2001, Eschrig2006} have been proposed to reproduce hourglass-shaped excitations. However, owing to the insufficient experimental resolution and the intensity of the neutron beam in earlier measurements, the determination of adequate models from the observed fuzzy spectra was difficult~\cite{Sato2013}. Furthermore, the gapped excitations in La$_{2-x}$Sr$_{x}$CuO$_4$ for 0.16 $\leq x \leq$ 0.21 exhibited other characteristics of spin fluctuations, i.e., the local spin susceptibility ($\chi''$) as a function of $\hbar\omega$ shows the peak-dip-hump structure with the peak- and hump-energies of $\sim$20~meV and $\sim$50~meV, respectively~\cite{Vignolle2007, Li2018}. This structured $\chi''$ suggests the existence of two characteristic energies for the spin fluctuations. Hence, the elucidation of the origin of such excitations would be essential to bridge the understanding of spin dynamics in La214 having a lower optimal $T_{\rm c}$ and that in a high-$T_{\rm c}$ cuprate family. 

In this study, we performed INS measurements on a LSCO to determine the detailed magnetic excitations in the $\hbar\omega$ range below $\sim$60 meV encompassing $E_{\rm cross}$ $\sim$35--40 meV of $x$ = 0.10 -- 0.16~\cite{Kofu2008, Vignolle2007}. To clarify the common feature of excitations in the SC phase, underdoped (UD) LSCOs with $x$ = 0.10 ($T_{\rm c}$ = 29 K) and OP ones at $x$ = 0.16 ($T_{\rm c}$ = 38 K) were measured. 
The former exhibited an IC magnetic order below 20 K \cite{Kimura1999}, whereas the latter indicated a clear spin gap in the excitations with the gap energy of $\sim$ 4 meV~\cite{Lake1999, Yamada1995, Mason1993}. Analysis based on the two-component picture, in which IC and C components were assumed to coexist in the wide $\hbar\omega$ range, yielded a gapless C component (spin-wave-like excitation) and an uprightly standing IC component, which faded away around $E_{\rm cross}$ with increasing $\hbar\omega$. 
The IC component weakened gradually toward the pseudo-gap (PG) temperature ($T$*) upon warming, whereas the C component was robust against temperature. These results suggest the itinerant spin nature in the IC component and the residual local antiferromagnetic correlations of the parent La$_2$CuO$_4$ in the C component, as reported for OP Y123~\cite{Reznik2008} and OP Hg1202~\cite{Chan2016} with higher $T_{\rm c}$. 

\section{Experimental Details}{\label{sec:experimental details}}

The single crystals of LSCO ($x$ = 0.10 and 0.16) for the INS measurements were grown by the traveling-solvent floating-zone method. The typical size of the crystal was 8 mm in diameter and 40 mm in length. Five single crystals were co-aligned with a total mosaic less than 1.5$^\circ$, and the entire mass weighed 71 and 55 g for $x = 0.10$ and 0.16, respectively. We confirmed a sharp superconducting transition at the onset of $T_{\rm c}$ = 29 and 38 K for $x$ = 0.10 and 0.16, respectively, by magnetic susceptibility measurements using an SC quantum interference device magnetometer (Quantum Design MPMS). The evaluated $T_{\rm c}$ was consistent with that reported previously~\cite{Takagi1989}. Furthermore, we directly measured the actual composition by inductively coupled plasma atomic emission spectroscopy measurements and confirmed that it was approximately the same as the nominal values.

\begin{table}[t]
\begin{center}
\small
\caption{Experimental conditions (Sr concentration, incident neutron energy, chopper frequency, energy-resolution (FWHM) at 40 meV (20 meV), momentum-resolution (FWHM) at 40 meV (20 meV) and (0.5, 0.5) and measured temperature)
 }
\begin{tabular}{p{4mm}|p{13mm}|p{9mm}|p{14mm}|p{13mm}|p{8mm}}
 $x$ & $E_i$ (meV) & $f$ (Hz) & $\Delta$$E$ (meV) & $\Delta$$h$ (r.l.u.) & $T$ (K)\\
 \hline
\footnotesize{0.10}  & \footnotesize{60.4 (30.1)}& \footnotesize{250} & \footnotesize{1.3 (0.6)} & \footnotesize{0.05 (0.04)}& \footnotesize{5, 250}\\
\footnotesize{0.16}  & \footnotesize{71.1 (33.7)}& \footnotesize{250} & \footnotesize{2.2 (0.8)} & \footnotesize{0.06 (0.04)}& \footnotesize{5, 300}\\ 
\end{tabular}
\end{center}
\label{table001}
\end{table}%

\begin{figure}[t]
\begin{center}
	\includegraphics[width=64mm]{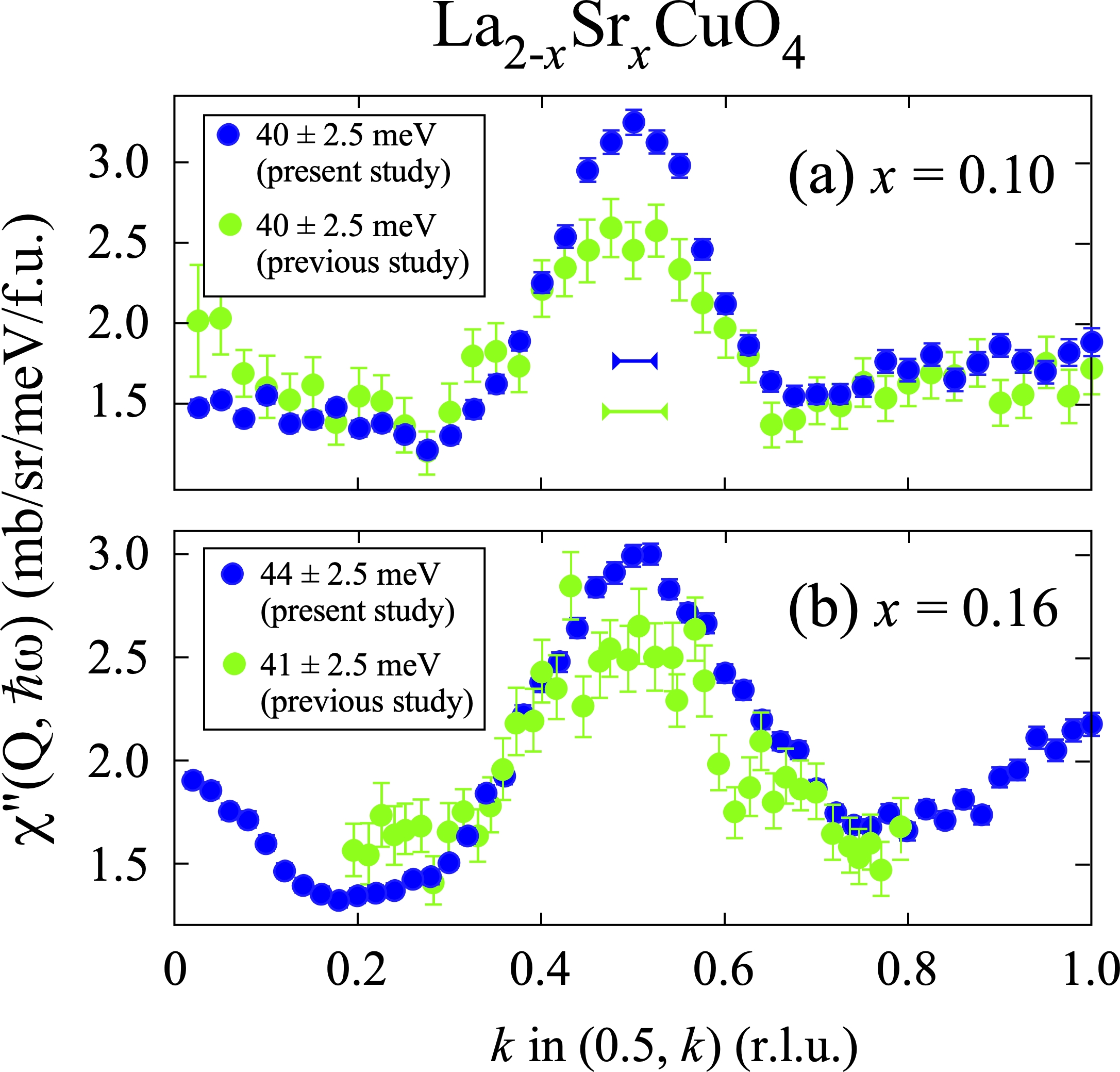}
\caption{(Color online) Constant-$\hbar\omega$ slices of $\chi"({\bf Q}, \hbar\omega)$ at around $E_{\rm corss}$ in La$_{2-x}$Sr$_x$CuO$_4$ (a) $x$ = 0.10 and (b) 0.16. The spectra are compared with the previously obtained ones by our measurement to show the improvement in data quality~\cite{Sato2013}. The spectra obtained in the previous study were shifted downward for a clear comparison. The horizontal bars represent the momentum resolution. 
 }
\label{fig0}
\end{center}
\end{figure}

The INS experiments were performed on the Fermi chopper spectrometer 4-SEASONS installed at BL01 of the Materials and Life Science Experimental Facility in J-PARC~\cite{Kajimoto2011}. Using the repetition-rate multiplication method \cite{Nakamura2009, Inamura2013}, we efficiently collected the magnetic scattering intensity by utilizing a series of incident energies ($E_i$) in a single measurement. 
We used $E_i$ = 30.1 and 60.4 meV for $x$ = 0.10 ($E_i$ = 33.7 and 71.1 meV for $x$ = 0.16) and the chopper frequency ($f$) of 250 Hz to investigate the magnetic excitation below $\sim$60 meV covering $E_{\rm cross}$. The latter $E_i$ with $f$ = 250 Hz provided an energy-resolution ($\Delta$$E$) of 1.3 meV (2.2 meV) and a momentum resolution along the tetragonal direction ($\Delta$$h$) of 0.05 (0.06) at $\hbar\omega$ = 40 meV and at {\bf Q} = (0.5, 0.5)~\cite{Kajimoto2018}. The experimental conditions for both samples are summarized in Table. I. Compared to the previous measurements~\cite{Sato2013}, the signal-to-noise (S/N) ratio improved from 0.5 to 1.3. The statistical accuracy of intensity was increased with the statistical error to the signal ration from 0.069 to 0.015 within a reasonable measurement time (36--52 h at each temperature). The representative spectra at $\sim$$E_{\rm cross}$ in the present and previous measurements~\cite{Sato2013} are shown in Fig. \ref{fig0}. We shifted the spectra obtained in the previous measurements downward for a clear comparison. Moreover, we successfully obtained high-quality data in this study. In the figure, horizontal bars indicate the momentum resolution in the full width at half maximum (FWHM) along the $k$-direction in (0.5, $k$).

%
%

%
\begin{figure}[t]
\begin{center}
	\includegraphics[width=\linewidth]{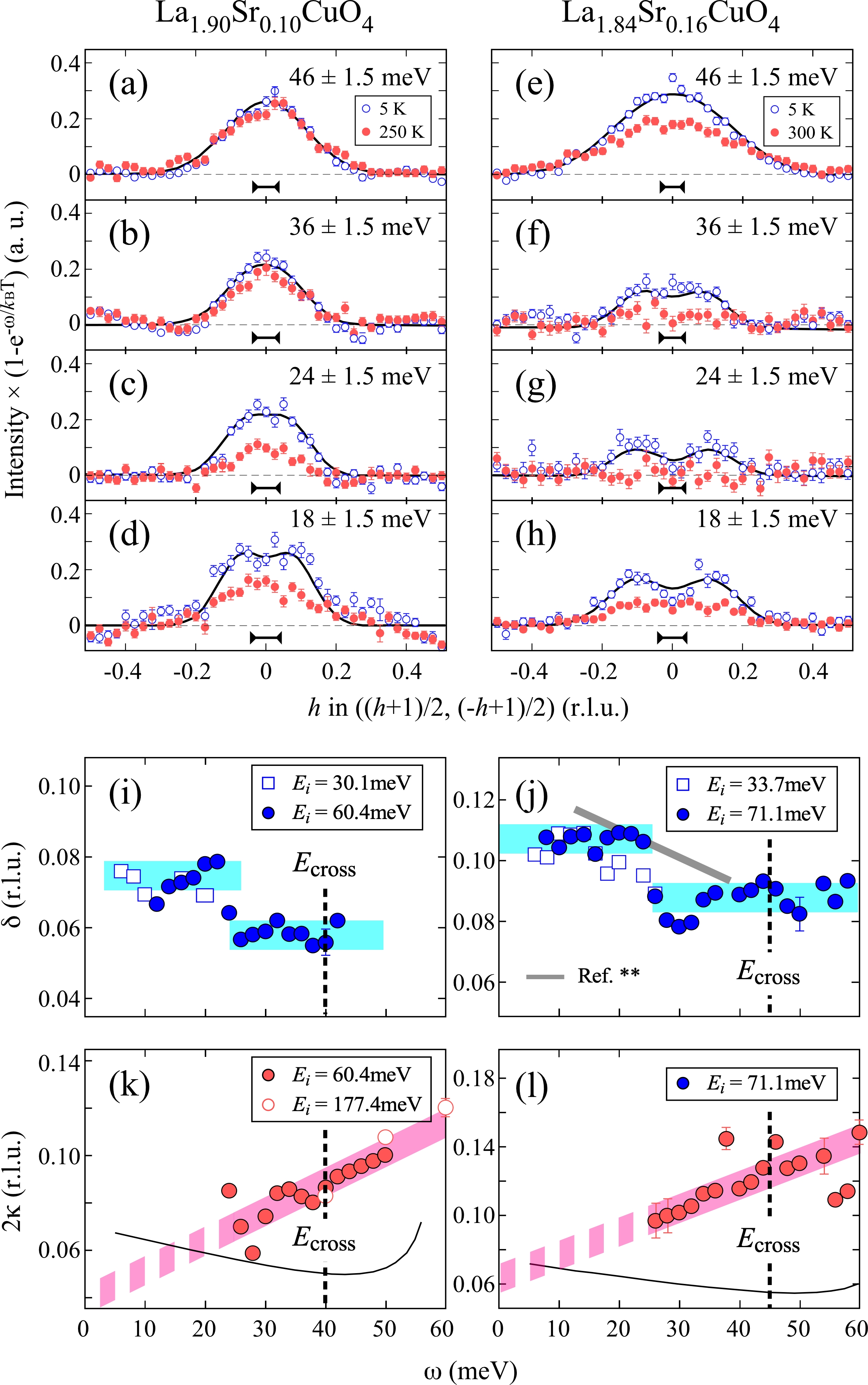}
\caption{(Color online) Constant-$\hbar\omega$ slices of $\chi"({\bf Q}, \hbar\omega)$ along [1, -1] direction at several energies in La$_{2-x}$Sr$_x$CuO$_4$ with (a)--(d) $x$ = 0.10 and (e)--(h) $x$ = 0.16. The horizontal bars correspond to the momentum resolution. Fitting results of incommensurability $\delta$ and peak-width 2$\kappa_{\rm C}$ (FWHM) of C component obtained by assuming only IC and C component respectively for (i), (k) $x$ = 0.10 and (j)--(l) $x$ = 0.16. (See the text.) The thick bars provide visual guidance. In Fig. \ref{fig1} (j), the gray line represents the result extracted from Ref. 4. 
The solid lines in Figs. \ref{fig1} (k) and (l) represent the momentum resolution as a function of $\hbar\omega$. 
 }
\label{fig1}
\end{center}
\end{figure}
%

The differential cross-section for a magnetic inelastic scattering is expressed as follows:
\begin{eqnarray}
\frac{d^2\sigma}{d\Omega dE} &=& \frac{2(\gamma r_{e})^2}{\pi g^2\mu_{B}^2}
\frac{k_{f}}{k_{i}} S({\bf Q},\hbar\omega)|f({\bf Q})|^2 \label{eqn:sq}. 
\end{eqnarray}
Here, $(\gamma r_{e})^2 = $0.2905 [barn], $g = 2$, and $\mu_{B}$ is the Bohr magneton. $k_i$ and $k_f$ represent the initial and final neutron wave vectors, respectively. 
The $S({\bf Q},\hbar\omega)$ ($f({\bf Q})$) is a dynamical structure factor (a magnetic form factor), which corresponds to the Fourier transform of the spin--spin correlation function of the target system (Fourier transform of the spin density distribution in the real space). 
$S({\bf Q},\hbar\omega)$ is related to the dynamical spin susceptibility $\chi"({\bf Q},\hbar\omega)$ through fluctuation--dissipation theorem,
\begin{eqnarray}
S({\bf Q},\hbar\omega)&=&\chi''({\bf Q},\hbar\omega) \frac{1}{1-e^{-\frac{\hbar\omega}{k_{B}T}}} \label{eqn:chiq}.
\end{eqnarray}

\begin{figure}[t]
 \begin{center}
	\includegraphics[width=\linewidth]{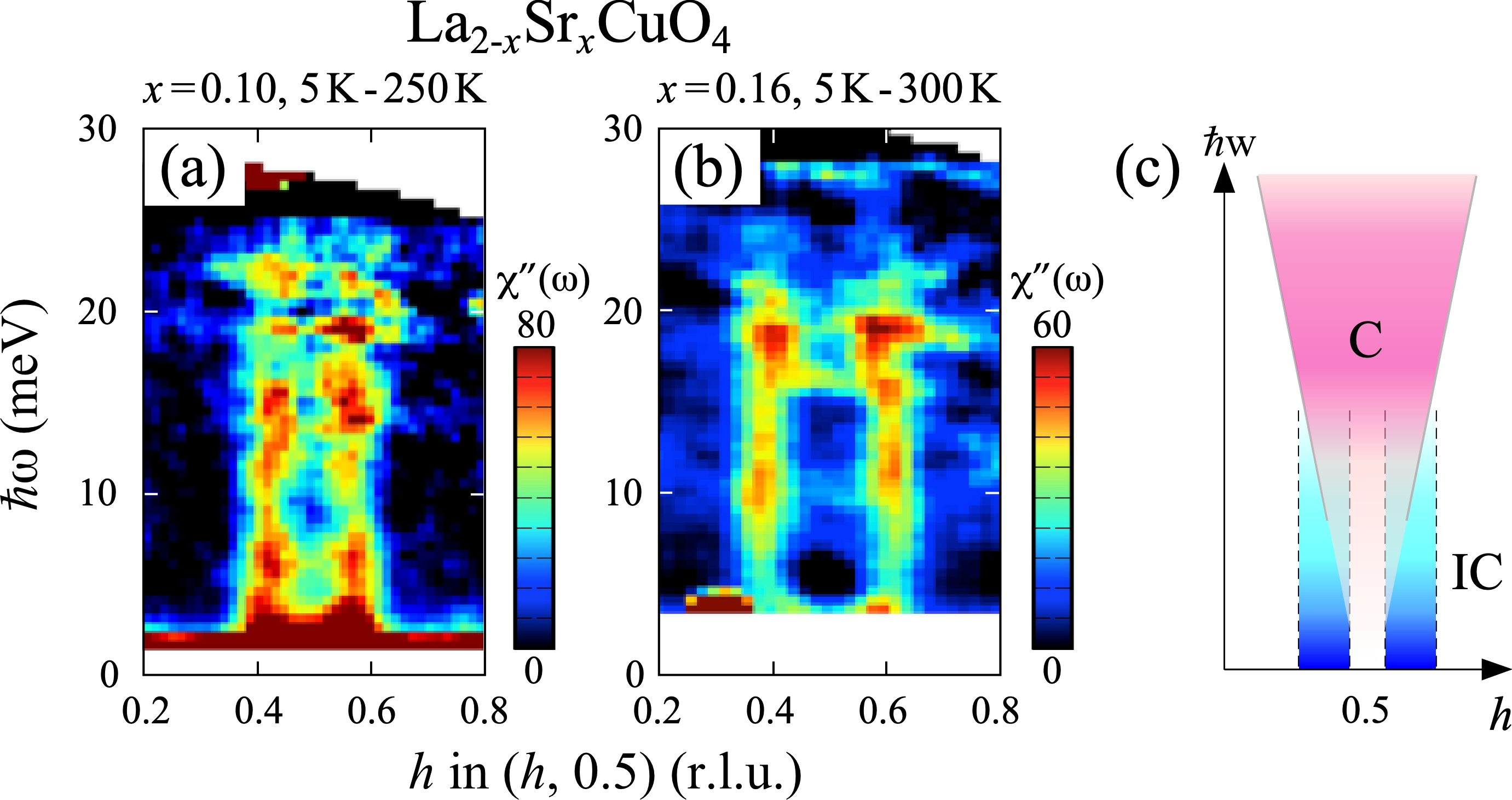}
  \caption{(Color online) Magnetic excitation spectrum in La$_{2-x}$Sr$_x$CuO$_4$ with (a) $x$ = 0.10 and (b) $x$ = 0.16. The spectrum were obtained by subtracting $\chi"({\bf Q}, \hbar\omega)$ at 250 K (300 K) from that at 5 K for $x$ = 0.10 ($x$ = 0.16). Color bars indicate the residual dynamical spin susceptibility in units of $\mu_{B}^{2} sr^{-1} eV^{-1} Cu^{-1}$. (c) Schematics of assumed spectra for the analysis, where IC and C components coexist in the energy and momentum spaces. (See the text.)}
  \label{fig2}
 \end{center}
\end{figure}
%

%
\begin{figure}[t]
\begin{center}
	\includegraphics[width=\linewidth]{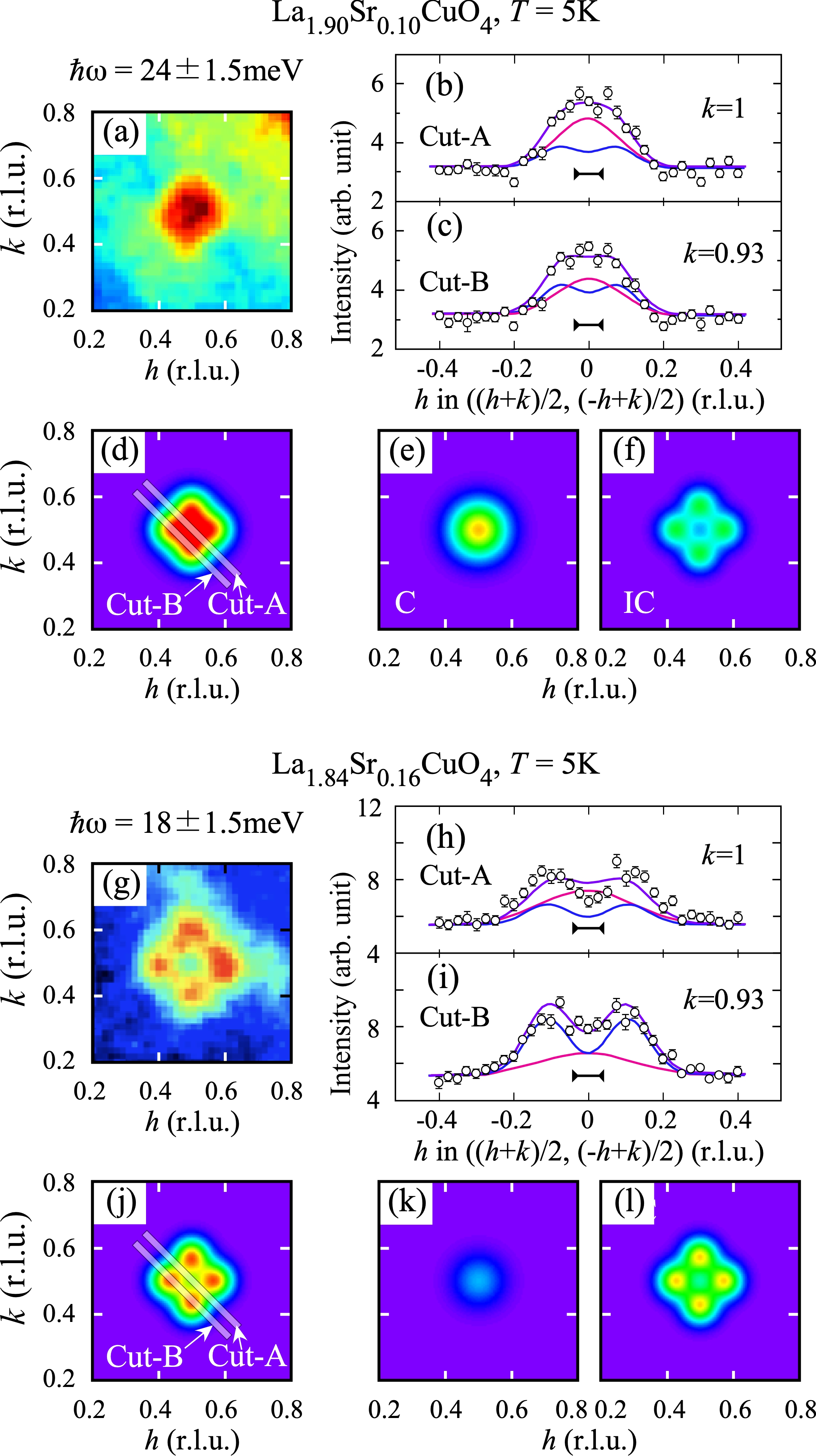}
\caption{(Color online) The measured $\chi''$({{\bf Q}, $\hbar\omega$}) for La$_{2-x}$Sr$_x$CuO$_4$ with (a) $x$ = 0.10 at 24 meV and (b) $x$ = 0.16 at 18 meV. Corresponding spectra in (b), (c) $x$ = 0.10 and (h), (i) $x$ = 0.16 sliced along the directions shown in Fig. \ref{fig3} (d) and (j). In the figure, the purple line represents the best-fitting results by two-component analysis, and blue and red lines represent the IC and C components, respectively. The reproduced $\chi''$({{\bf Q}, $\hbar\omega$}) for (d) ((j)) the superposition of the two components and (e), (f) ((k)--(l)) the individual component for $x$ = 0.10 ($x$ = 0.16) with the parameters obtained by fitting. 
}
\label{fig3}
\end{center}
\end{figure}
%

The scattering intensity was converted into $S$({{\bf Q}, $\hbar\omega$) in the absolute units using the magnetic form factor $f$({\bf Q}) for the Cu 3$d$$_{x^2-y^2}$ orbital and the incoherent cross-section of the samples. In this study, we analyzed $S$({{\bf Q}, $\hbar\omega$) and evaluated the momentum-integrated $\chi''$({{\bf Q}, $\hbar\omega$) ($\chi''$($\hbar\omega$)) to discuss the two components in the excitation spectra. The momentum transfer ($h$, $k$, $l$) was labeled in units of reciprocal lattice with $a$* = $b$* = 1.661 (1.663) ${\rm \AA}$$^{-1}$ and $c$* = 0.4758 (0.4748) ${\rm \AA}$$^{-1}$ in the tetragonal notation for $x = $0.10 (0.16). 

\section{Experimental Results}{\label{sec:experimental results}}

In Figs. \ref{fig1} (a)--(h), the constant-$\hbar\omega$ spectra are shown along the diagonal [1, -1] direction through (0.5, 0.5). In the spectrum along this direction, the phonon intensity can be eliminated; therefore, the detailed structure of the magnetic signal can be investigated. The horizontal bars in the figure represent the momentum resolution. The thermal factor was corrected for the spectrum at each $\hbar\omega$ after subtracting the constant background. At 5 K, the IC structure can be observed for $\hbar\omega$ $\lesssim$ 23 meV (27 meV) in $x$ = 0.10 (0.16). The IC peaks vanished or became a broad single peak at high temperatures, whereas the C peak at the high-$\hbar\omega$ region indicated a negligible or weak temperature dependence. This thermal evolution of excitation spectra was observed in L$_{1.875}$aBa$_{0.125}$CuO$_4$~\cite{Xu2007}. Therefore, the IC structure is characteristic of low-temperature magnetic excitations. 

Following earlier studies, we first analyzed the spectra at 5 K by considering four equivalent isotropic IC peaks at (0.5$\pm\delta$,0.5)/(0.5,0.5$\pm\delta$)~\cite{Tranquada2004, Vignolle2007, Lipscombe2009}. In the analysis, two-dimensional Gaussian functions were used to describe the IC peaks, and least-squares fitting was performed on $\chi''({\bf Q},\hbar\omega)$. Here, the fitting parameters are the peak intensity, incommensurability($\delta$), and the width of the IC peaks in FWHM (2$\kappa_{\rm IC}$). The solid lines in Figs. \ref{fig1} (a)--(h) are the fitting results. For $x$ = 0.10, the superposition of four IC peaks reasonably reproduced a broad single peak above $\sim$23 meV and a signal below $\sim$23 meV. 
However, as shown in Fig. \ref{fig1} (i), the evaluated $\delta$ as a function of $\hbar\omega$ shows an anomalous kink at $\sim$23 meV; the constant values of $\delta$ differ at above and below $\sim$23 meV. 

The same trend was observed for $x$ = 0.16. (See Fig. \ref{fig1} (j).) This result differs from the gradual decrease in $\delta$ with increasing $\hbar\omega$, which has been reported in earlier studies. In Fig. \ref{fig1} (j), the $\hbar\omega$-dependence of $\delta$ reported for $x$ = 0.16 is represented by a gray solid line~\cite{Vignolle2007}. Hence, the high-quality data obtained with better experimental resolution and S/N ratio yielded a new aspect in magnetic excitations. This discontinuous behavior is difficult to understand based on either the stripe or nesting model. The anomaly is likely owing to the analysis, which assumes only the IC component and provides strong evidence for the coexistence of C and IC components.

%
 \begin{figure}[t]
\begin{center}
	\includegraphics[width=\linewidth]{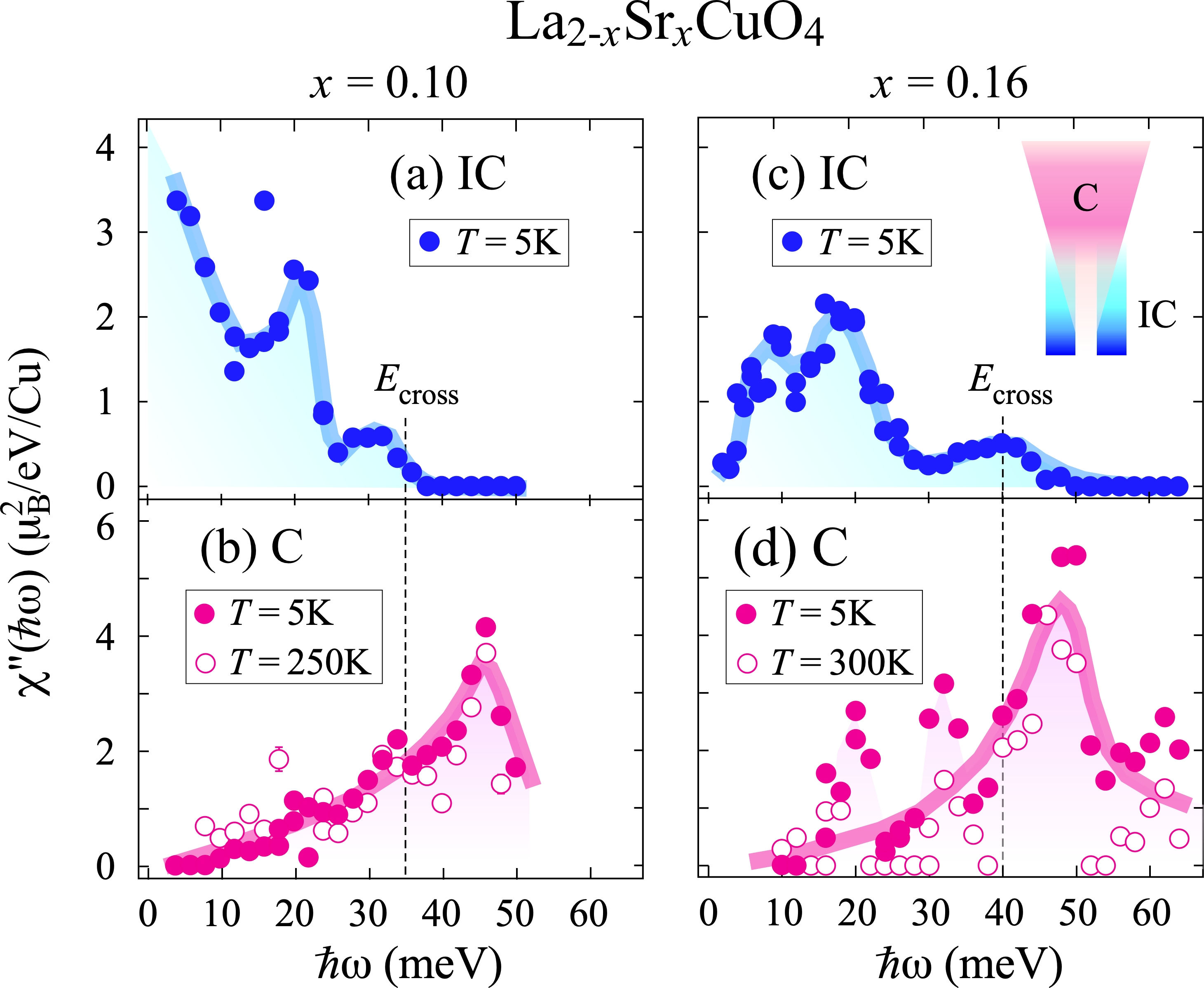}
\caption{(Color online) Local spin susceptibility for decomposed (a), (c) IC (closed square) and (b), (d) C components (closed circle) in $x$ = 0.10 and 0.16. 
The solid lines provide visual guidance. The open circle indicates the C component at high temperatures. Inset figure in Fig. \ref{fig4} (c) represents the schematics of spectra used for the analysis, where IC and C components are assumed to coexist in the energy and momentum spaces. }
\label{fig4}
\end{center}
\end{figure}
%

To obtain further insights into the coexistence of the two components, the high-temperature spectra were subtracted from the low-temperature ones. Figures \ref{fig2} (a) and (b) show the residual $\chi$"({{\bf Q}, $\hbar\omega$) for $x$ = 0.10 and 0.16, respectively. The vertical and horizontal axes correspond to $\hbar\omega$ and $h$ in ($h$, 0.5), respectively. The magnetic excitations with upright IC structure are revealed, i.e., the $\delta$ and $\kappa$ are almost constant with respect to $\hbar\omega$. Subsequently, we reanalyzed the low-temperature spectra by assuming the coexistence of C and vertically standing IC components. In the fitting process, we used fixed values for $\delta$ and 2$\kappa_{\rm IC}$ of 0.076 (r.l.u.) and 0.05 (r.l.u.) for $x$ = 0.10 (0.107 (r.l.u.) and 0.05 (r.l.u.) for $x$ = 0.16) in the entire $\hbar\omega$ range, which are the average values for $\hbar\omega$ $\leq$ $E_{\rm kink}$ obtained from the analysis above by considering only the IC component. This assumption is consistent with the experimental fact that $\delta$ and $\kappa_{\rm IC}$ are almost $\hbar\omega$-independent below $\sim$50 meV in the overdoped (OD) LSCO with $x$ = 0.22, in which the IC structure was more apparent owing to the strong suppression of the C component around $\sim$50 meV~\cite{Lipscombe2007, Wakimoto2004}. Furthermore, we assumed a linear relation between the peak-width for the C component (2$\kappa_{\rm C}$) and $\hbar\omega$. For the evaluation of 2$\kappa_{\rm C}$, we analyzed the spectra above $E_{\rm kink}$ by assuming an isotropic C peak centered at (0.5, 0.5). The spectra were fitted by a two-dimensional Gaussian function, and the fitted results of 2$\kappa_{\rm C}$ for $x$ = 0.10 and 0.16 are shown in Figs. \ref{fig1} (k) and (l), respectively. The additional data obtained with $E_{\rm i}$ = 177.4 meV is plotted for $x$ = 0.10. We estimated 2$\kappa_{\rm C}$ for the low-$\hbar\omega$ region from the extrapolation of the results at a high-$\hbar\omega$ region and then used the evaluated values for further analysis. Hence, the fitting parameters were the intensity of the C and IC components. In the previous two-component analysis, the change in magnetic components crossing $E_{\rm cross}$ was assumed~\cite{Tranquada2004, Vignolle2007, Lipscombe2009}. By contrast, the present modified two-component picture considered the coexistence of IC and C components at the constant $\hbar\omega$. The representative fitting results for $x$ = 0.10 and 0.16 are depicted in Figs. \ref{fig3} (b), (c), (h), and (i). All the spectra, including the flat-top features, can be well reproduced by the superposition of two components. In the figure, the purple line is the best-fitting result. The red and blue lines represent the C and IC components, respectively. The intensity distribution of each component as well as the total intensity in the ($h$, $k$) plane are shown in Figs. \ref{fig3} (e), (f), and (d) for $x$ = 0.10, respectively (Figs. \ref{fig3} (k), (l), and (j) for $x$ = 0.16, respectively). The overall shape of the total intensity agreed reasonably well with the observed spectra. (Figs. \ref{fig3} (a) and (g).) 

Next, we present the resultant $\hbar\omega$-dependence of decomposed $\chi$" for each component as one of the main results of this study. Figures \ref{fig4} (a) and (b) ((c) and (d)) represent $\chi$" for the IC and C components for $x$ = 0.10 (0.16), respectively. The result of the analysis indicates the existence of the IC component at 5 K below $\sim$40 meV, which is comparable to $E_{\rm cross}$. The $\chi$" for IC component in UD $x$ = 0.10 decreased gradually with increasing $\hbar\omega$, whereas that in OP $x$ = 0.16 exhibited a gap structure with the gap energy of $\sim$5 meV. These features were consistent with the results obtained using triple-axis spectrometers~\cite{Lake1999, Yamada1995, Mason1993}. Meanwhile, the intensity of the C component increased from the low-$\hbar\omega$ region with increasing $\hbar\omega$ and strengthened at a higher $\hbar\omega$. At high temperatures, no obvious IC peaks were observed at all measured $\hbar\omega$. Therefore, we analyzed the spectra at 250 and 300 K by the C component only. As shown in Fig. \ref{fig4}, the C component remained at an intensity comparable to that at 5 K. Hence, these results suggest the individual thermal evolution of the two components. 

To confirm the different temperature dependences of $\chi$" for the two components, the analysis above based on the two-component picture was applied to the spectra for $\hbar\omega$ = 10 meV at several temperatures. For evaluation, we mainly used the data of $x$ = 0.10 obtained previously using the same spectrometer\cite{Sato2014}. Although the measurements were performed with a relatively poor experimental resolution, the decomposition of $\chi$ should be applicable to the spectrum in the low-$\hbar\omega$ region, as the spectral change from IC to C upon warming was observed. In Figs. \ref{fig5} (a)--(e), the spectra at 10 meV slice along the diagonal direction are shown. The IC peaks damped gradually upon warming and transformed into a C peak at high temperatures. As shown by the blue and red lines, which represent the best-fitting of the IC and C components, respectively, the relative intensity of the C component increased with  temperature. Fig. \ref{fig5} (f) shows the temperature dependence of $\chi$ for the IC component ($\chi_{\rm IC}$) and that for the C component ($\chi_{\rm C}$) at 10 meV. $\chi$ was normalized such that the total intensity at 5 K became 1. More importantly, the IC component weakened toward $T$*, which was determined by the angle-resolved photoemission spectroscopy measurement~\cite{Yoshida2009}. By contrast, the C component strengthened slightly upon heating in the entire temperature range. We plotted the intensity for $\hbar\omega$ = 45 meV, as shown in Fig. \ref{fig5} (g), where only the commensurate peak was observed. The intensity was approximately constant at low temperatures and began to decrease at $\sim$$T$* upon warming. This behavior was consistent with the observation for the LSCO with $x$ = 0.075, which indicated the modulation of the spin-wave-like high-$\hbar\omega$ branch into a broad ridge centered at (0.5, 0.5) upon warming by crossing $T$*. As the origin of modulation was related to the itinerant carriers, the similarity suggested that the C component was part of the spin-wave-like excitations appearing in the high-$\hbar\omega$ region and reflected the localized spin nature below $T$*~\cite{Matsuura2017}. 

%
\begin{figure}[t]
\begin{center}
	\includegraphics[width=\linewidth]{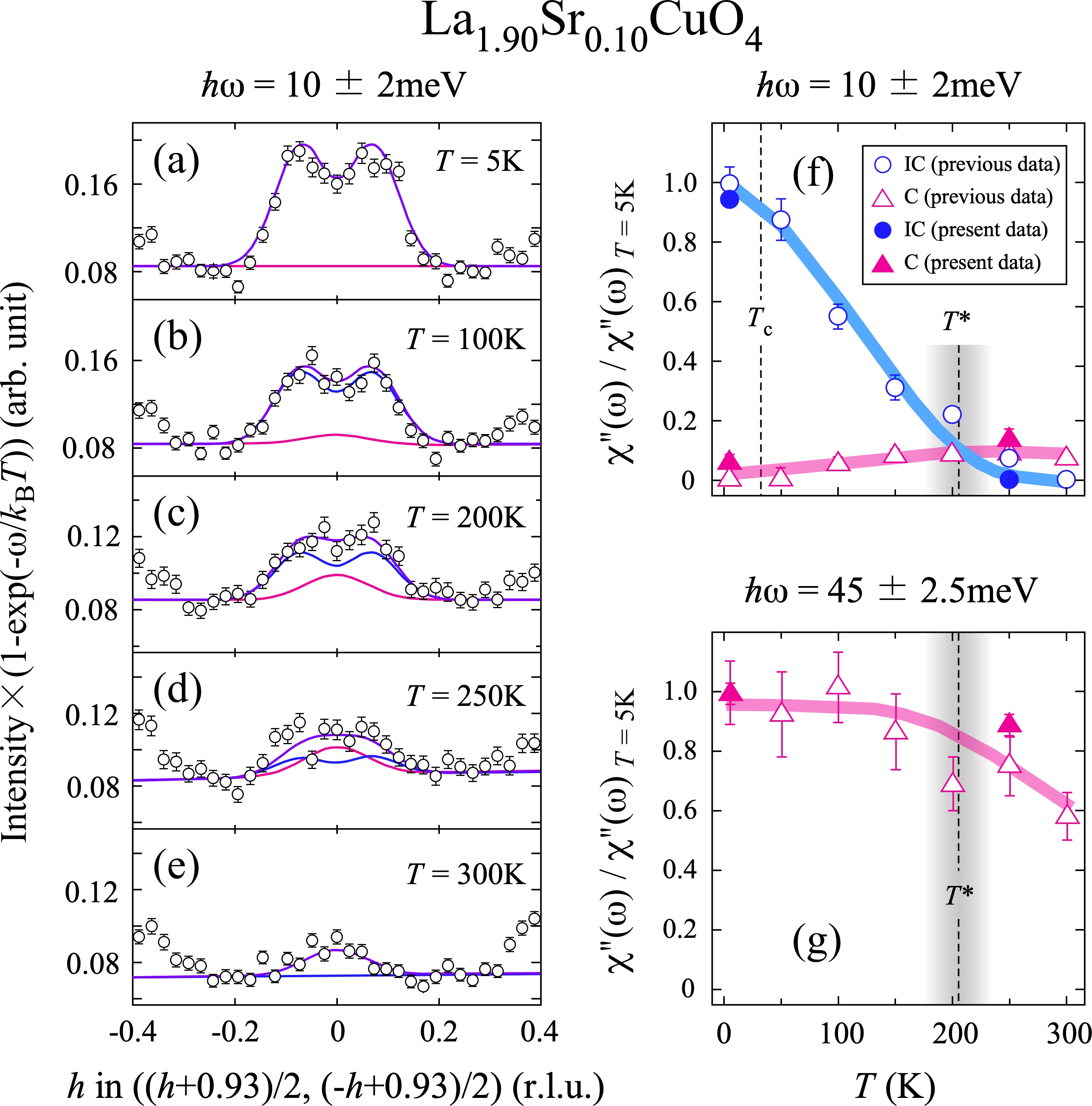}
\caption{(Color online) (a)--(e) Magnetic excitation along [1, -1] direction in $x = 0.10$ at different temperatures. The purple line is the best-fitting result by two-component analysis; blue and red lines represent the IC and C components, respectively. The temperature dependence of local spin susceptibility of each component at (f) 10 meV and (g) 45 meV. The circle and triangle symbols correspond to the IC and C components, respectively. The closed and open symbols represent the results of the analysis for the present and previously measured data, respectively \cite{Sato2014}. The thick lines provide visual guidance.
}
\label{fig5}
\end{center}
\end{figure}
%

\section{Discussion}

In this study, we analyzed the high-quality data based on the two-component picture and obtained new findings regarding magnetic excitations. First, both decomposed $\chi _{\rm IC}$ and $\chi _{\rm C}$ in magnetically ordered UD $x$ = 0.10 and OP $x$ = 0.16 showed similar $\hbar\omega$-dependence, except for the difference in the low-temperature $\chi_{\rm IC}$ below $\sim$ 5 meV. Therefore, the characteristics of magnetic excitations in the two samples were the same, and the origin of the entire excitations was considered to be identical. The results of two-component analysis further suggested a new perspective on the hourglass excitation. Because 2$\kappa_{\rm C}$ was narrower than the distance between a pair of IC peaks (constant 2$\delta$) in the low-$\hbar\omega$ region, the superposition of the two components with a particular intensity balance formed an hourglass-shaped excitation. Therefore, $E_{\rm cross}$ corresponded to $\hbar\omega$, where the C component dominated with increasing $\hbar\omega$. Although the extension mechanism of the IC structure in the low-$\hbar\omega$ region to the high-$\hbar\omega$ region remain unclear, the upright structure was a solid feature of the IC component. The uprightly standing structure and the thermal evolution of the IC component were consistent with the theoretical study by Norman that indicated the appearance of particle--hole pair excitations in PG and SC states~\cite{Norman2001}. Therefore, we speculate that most of the low-$\hbar\omega$ components except for low-$\hbar\omega$, where the spectra between UD $x$ = 0.10 and PD $x$ = 0.16 differed, reflects the response of the itinerant electron spins and the spectra related to the Fermi surface topology. 

Furthermore, we discovered that the C component exhibited a finite intensity even in the low-$\hbar\omega$ region at both high and low temperatures. The $\chi$" in $x$ = 0.10 showed no clear gap structure and was extrapolated to zero value at the elastic position ($\hbar\omega$ = 0 meV). Hence, the gapless C component extended up to the high $\hbar\omega$ region exceeding 200 meV~\cite{Sato2013, Kofu2008, Meyers2017}. The existence of gapless excitations implied that spin--wave-like excitations remained in the SC phase of the LSCO with the strong suppression of the intensity at low-$\hbar\omega$ owing to carrier doping. Matsuura et al. reported that the C component can be induced in a low $\hbar\omega$ region by Ni doping, and that the induced C component coexisted with the originally existing IC component in La$_{1.85}$Sr$_{0.15}$Cu$_{1-y}$Ni$_y$O$_4$~\cite{Matsuura2012}. The intensity appeared gradually from the higher $\hbar\omega$ region below $E_{\rm cross}$ $\sim$ 40 meV of the pristine LSCO owing to Ni substitution and increased in the lower-$\hbar\omega$ region with the amount of Ni. This Ni-induced structural change was interpreted as the extension of the upward branch of the hourglass excitation to the lower energies with a reduce in the minimum energy of the branch (C component) from $E_{\rm cross}$ for $y$ = 0 to 0 meV for $y$ = 0.04~\cite{Matsuura2012, Kofu2005}. In the present two-component picture, such an increase in intensity at the C position was assumed to be due to the enhancement of the inherently existing C component. As mentioned above, the particular $\hbar\omega$-dependence of the $\chi_{\rm IC}$ and $\chi_{\rm C}$ components determines $E_{\rm cross}$. Hence, the enhancement of the C component at the lower $\hbar\omega$ with increasing $y$ resulted in the reduction in $E_{\rm cross}$. In any case, the magnetic excitations in La$_{1.85}$Sr$_{0.15}$Cu$_{1-y}$Ni$_y$O$_4$ supported the presence of low-energy C components and the coexistence of C and IC components at a constant $\hbar\omega$. 

%
\begin{figure}[t]
\begin{center}
	\includegraphics[width=\linewidth]{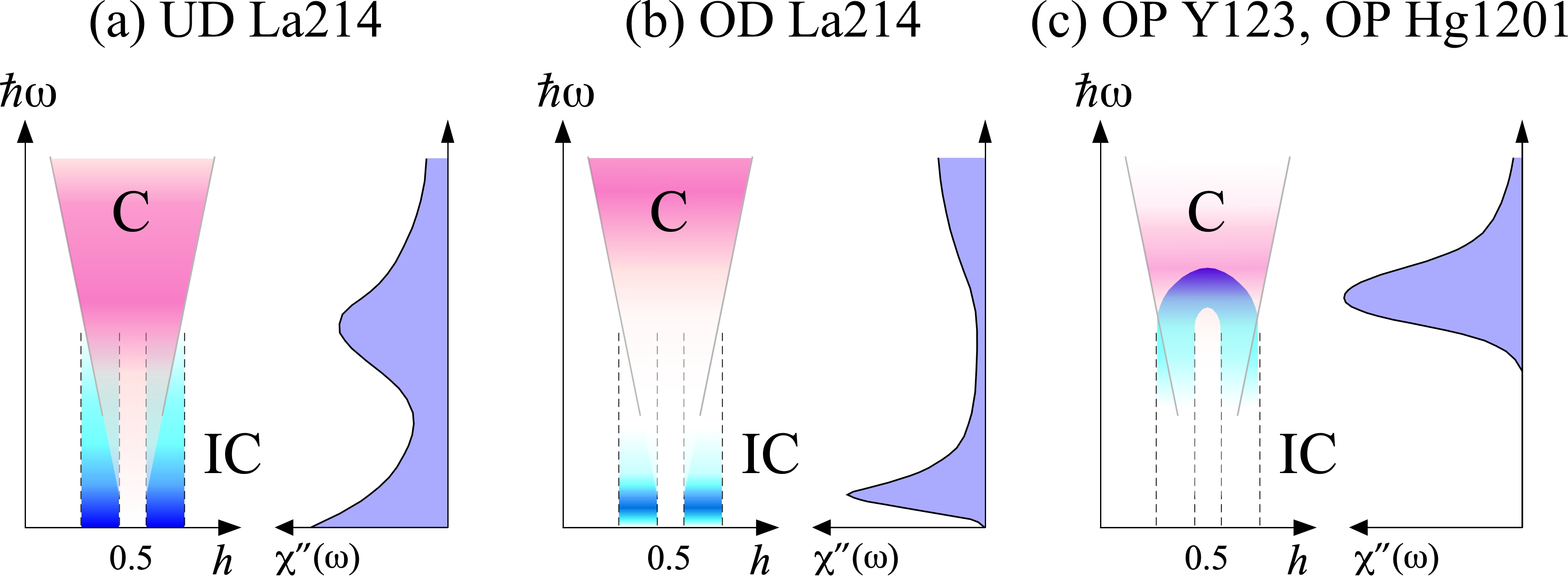}
\caption{(Color online) Schematics of the decomposed magnetic excitations, and the local spin susceptibilities of (a) UD La124, OD La124, and (c) OP Y123 and Ha1201, which were inferred from the two-component picture.
}
\label{fig6}
\end{center}
\end{figure}
%

Finally, we present the difference in and the unified view of magnetic excitations in a few cuprate oxides. 
For the IC component, the low-$T_{\rm c}$ La214 exhibited a large intensity in the low-$\hbar\omega$ region, whereas the high-$T_{\rm c}$ Y213 and Hg1201 showed a gap structure with the gap energy of $\sim$30 -- 40 meV. Additionally, our analysis revealed a difference in the C component between the two families, i.e., the C component in the present LSCO exhibited a gapless feature, whereas OP YBCO and OP Hg1201 showed no well-defined C components below the resonance energy near $E_{\rm cross}$. The absence of the intensity of the C component at the lower $\hbar\omega$ region was also observed in the OD LSCO; the hump structure in $\chi$ at $\sim$50 meV present in the UD LSCO was suppressed, and the magnetic intensity emerged above $\sim$100 meV with increasing $\hbar\omega$~\cite{Lipscombe2007}. Considering a nearly $\hbar\omega$-independent $\chi$" in La$_2$CuO$_4$~\cite{Hayden1991}, we speculate that the spin--wave-like gapless excitation loses intensity in the lower-$\hbar\omega$ region upon doping~\cite{Fujita2012}. 
In this context, the low-$\hbar\omega$ C component of the OP high-$T_{\rm c}$ family is significantly damped owing to the prominent effect from carrier doping. The difference in the C component (spin--wave-like gapless excitation) between the OP low-$T_{\rm c}$ La214 and  OP high-$T_{\rm c}$ family suggests the differing natures of the AF correlations of local moments between the OP sample of the two families. 
The schematic excitation spectra inferred from the two-component picture and the reported $\chi$" are shown in Fig. \ref{fig6} for UD~\cite{Vignolle2007} and OD La214~\cite{Lipscombe2009}, and OP Y123~\cite{Reznik2008} and Hg1201~\cite{Chan2016}. In the present study, we considered the two components separately; however, it may be necessary to examine the correlation between them. To understand the coexistence of the IC and C components in the reciprocal space, the corresponding real space image must be elucidated.

\section{Summary}

Magnetic excitations in La$_{2-x}$Sr$_x$CuO$_4$ with UD $x$ = 0.10 and OP $x$ = 0.16 were analyzed based on a two-component picture, which considered the coexistence of IC and C components. At low temperatures, the decomposed uprightly standing component faded away around $E_{\rm cross}$ upon increasing $\hbar\omega$, whereas the gapless C component (spin--wave-like excitation) increased with the intensity. Within the two-component picture, it was interpreted that a particular intensity balance of the two components formed an hourglass-shaped excitation, and $E_{\rm cross}$ corresponded to $\hbar\omega$, where the C component dominated with increasing $\hbar\omega$. Upon warming, the IC component disappeared near the pseudo-gap temperature, whereas the C component was robust against temperature. These results are qualitatively similar to those of the excitations in the high-$T_{\rm c}$ family of Y123 and Hg1201, which suggest the itinerant and localized spin nature of IC and C components, respectively. 

\section*{Acknowledgments}
We are grateful to M. Matsuura, M. Mori, T. Tohyama, O. Sushkov, and K. Yamada for helpful discussions. We also thank K. Nakajima for sharing the beamtime, and K. Iida and Y. Inamura for their help with measurements using the 4-SEASONS spectrometer. The neutron experiments at the J-PARC were performed under Project No. 2014P0201. M. F. is supported by the Grant-in-Aid for Scientific Research (A) (16H02125).

\end{document}